\documentclass{article}
\usepackage{spconf,amsmath,amssymb,graphicx,booktabs,microtype,array}
\usepackage{siunitx}
\usepackage[hidelinks]{hyperref}

\let\originalthebibliography\thebibliography%
\renewcommand{\thebibliography}[1]{%
  \originalthebibliography{#1}%
  \small
  \setlength{\baselineskip}{9pt}%
  \sloppy
  \setlength{\itemsep}{0pt}%
  \setlength{\parsep}{0pt}%
  \setlength{\topsep}{0pt}%
}
\title{ON A SEPARATE NOTE:
ROBUST SCORE-INFORMED NOTE SEPARATION WITH A TWO-STREAM TFC--TDF U-NET AND ADAPTIVE SET OWNERSHIP}

\name{\shortstack{Benjamin Shiue-Hal Chou$^{1}$ \qquad
Purvish Jajal$^{1}$ \qquad
Nicholas John Eliopoulos$^{1}$ \qquad
James C. Davis$^{1}$ \\
George K. Thiruvathukal$^{2}$ \qquad
Kristen Yeon-Ji Yun$^{1}$ \qquad
Hao-Wen Dong$^{3}$ \qquad
Yung-Hsiang Lu$^{1}$}}
\address{$^{1}$Purdue University \qquad
$^{2}$Loyola University Chicago \qquad
$^{3}$University of Michigan \\
\small\texttt{\{chou150,pjajal,neliopou,davisjam,yun98,yunglu\}@purdue.edu} \\
\small\texttt{gkt@cs.luc.edu} \quad \texttt{hwdong@umich.edu}}

\begin{document}
\maketitle

\begin{abstract}
Score-informed note separation seeks to extract the performed waveform of all
individual notes, often from a polyphonic recording. Existing deep learning systems generally only target instrument-level stems. We present, to our knowledge, the first deep learning approach to score-informed note separation, NoteSep. NoteSep extracts the queried notes by applying an extraction stage model, NoteGrab, once per note. Conditioned on pitch, onset, and offset, NoteGrab separates harmonic and percussive components in two U-Nets linked by bidirectional cross-attention; selective harmonic gating suppresses lower-octave interference while preserving percussive attacks. Finally, a joint separation stage applies Adaptive Set Ownership (ASO) to compare concurrent NoteGrab estimates and reallocate mixture energy. We curate SCNS-Train (25,729 mixtures and 743,920 targets) for training and SCNS-Eval (16 instruments, disjoint scores and libraries) for evaluation. On SCNS-Eval, NoteSep reaches a median SI-SDR of 7.39~dB, compared with 2.49~dB for our strongest baseline. See the demo page at
\href{https://benschou.com/notesep/}{benschou.com/notesep}.
\end{abstract}


\section{Introduction}

A wrong note can force a musician to repeat an otherwise successful take. A student or instructor may also
want to isolate a musical line or particular note to aid in learning.
Current music source separation (MSS) models usually operate at the
instrument-stem level, separating one instrument from the others. Universal separators could leverage audioe-aligned language embeddings, like CLAP~\cite{wu2023clap}, to specify semantically describable sources~\cite{huang2025zerosep}. However, notes from the same instrument
share similar timbre, so semantic embeddings provide little information for
distinguishing individual notes~\cite{shi2025samaudio}. Notes can also share
many frequencies or harmonic partials. Octave-related notes and perfect fifths share substantial harmonic content, while inharmonicity, vibrato, room acoustics, and percussive attacks make separation difficult~\cite{every2006synchronous,han2011overlapped}.

\begingroup\tolerance=9999\emergencystretch=4em\looseness=-1
Prior work on music source separation usually predicts instrument, vocal, or
some text prompt sources~\cite{manilow2020cerberus}. Score-informed systems use aligned notation to separate
orchestral instruments or choir sections~\cite{miron2016orchestral,
gover2020choral,tunturi2025scoremss}, but still return one waveform per voice
or part, such as tenor and bass, rather than an individual scored note.
\par\endgroup

\begingroup\tolerance=9999\emergencystretch=2em\looseness=-1
Past note-separation work relied mainly on harmonic comb filtering,
nonnegative matrix factorization (NMF), or score-informed template
decomposition~\cite{every2006synchronous,han2011overlapped,
driedger2013scoredecomp,ewert2012scoreinformednmf}. Their assumptions limit performance on octaves, inharmonic partials, and broadband attacks, as well as real world recordings where instrument timbre varies based on many factors. Other systems adapt harmonic and
inharmonic models to particular instruments or rely on post-human judgment for
better separation performance~\cite{itoyama2007harmonic,
celemony2024melodyne}. Consequently, prior systems either require substantial
instrument- or recording-specific adjustment, or trade off specialized systems for a
more general-purpose but low fidelity result.
\par\endgroup

\begingroup\tolerance=9999\emergencystretch=2em\looseness=-1
We present NoteGrab, a deep learning-based note extractor conditioned on an aligned
pitch, onset, and offset. Harmonic and percussive components of the mixture are fed through two TFC--TDF v3
U-Nets~\cite{kim2023tfctdf}. The two model streams process the harmonic--percussive source separation (HPSS) components of the mixture and exchange
intermediate features near the bottlenecks. A high-pass filter removes sub-fundamental harmonic
energy only when a lower-octave competitor overlaps, with the goal of removing octave collisions; the percussive stream
remains ungated to retain attacks and inharmonic detail. Each query produces
one note waveform. For complete separation of a mixture into a complete set of notes, our method Adaptive Set Ownership (ASO) compares all concurrent
estimates and reallocates shared mixture energy.
\par\endgroup

\begingroup\tolerance=9999\emergencystretch=4em\looseness=-1
We make two contributions. First, to the best of our knowledge, we introduce
the first deep learning-based method for score-informed note separation. 
Second, we curate SCNS-Train, a training set with 25,729 mixtures and
743,920 isolated note targets, and SCNS-Eval, a held-out evaluation set built
from recorded-note samples that do not appear in training. We evaluate NoteSep
across 16 instruments on SCNS-Eval, where NoteSep outperforms the
evaluated Score-Informed NMF~\cite{driedger2013scoredecomp} and Melodyne baselines. We further evaluate the
model on realistic recording scenarios by testing it's performance on real ensemble recordings from PHENICX-Anechoic and Bach10 by aggregating
the note estimates into instrument parts and comparing to prior specialized systems. We show Notesep achieves competitive performance.
Audio examples and an interactive note-level visualization are available on
the demo page.
Code and model weights will be released here\footnote{\href{https://github.com/ben2002chou/notesep}{github.com/ben2002chou/notesep}.}.
\par\endgroup

\section{Method}
\subsection{Task and system overview}

\begingroup\tolerance=9999\emergencystretch=4em\looseness=-1
A monaural mixture is modeled as \(x=\sum_i s_i\), where \(s_i\) is the
performed waveform of note \(i\). An aligned score event provides
\(q_i=(p_i,t_i^{\mathrm{on}},t_i^{\mathrm{off}})\): its pitch, onset, and
offset.
\par\endgroup

Given a local mixture window and \(q_i\), our score-conditioned
extractor, \textit{NoteGrab}, predicts \(\hat{s}_i\), an estimate of \(s_i\).
For a window of \(K\) queried events, \textit{NoteSep} applies NoteGrab to each
event and obtains \(K\) raw note estimates. Its joint ASO stage then
reallocates mixture energy among concurrent estimates.

\subsection{Inputs}
\begingroup\tolerance=9999\emergencystretch=4em\looseness=-1
For each query, we extract a mixture segment centered on the target note's onset
and offset. HPSS~\cite{fitzgerald2010hpss} separates this segment into harmonic and percussive
inputs.\footnote{We use the \texttt{librosa} implementation: \href{https://librosa.org/doc/latest/generated/librosa.decompose.hpss.html}{librosa HPSS documentation}.}
Then, two separate model streams process the input components. 
We compare two variants. In the ungated variant, both streams receive their
complete HPSS components. This leaves the complete target spectrum available
to the model, but it also exposes the harmonic stream to energy shared with
lower-octave notes.
\par\endgroup

\begingroup\tolerance=9999\emergencystretch=4em\looseness=-1
In the selective variant, we check the score for an overlapping note one or
two octaves lower. When one is present, we zero the harmonic input below
\(\max(30,f_0-60)\) Hz; otherwise, the stream receives the original harmonic
component. Thus, the gate targets the lower-octave ambiguity only when it
occurs. The percussive input remains ungated, leaving attacks, resonances, and
other inharmonic energy below \(f_0\) available to the model.
\par\endgroup

\begingroup\tolerance=9999\emergencystretch=4em\looseness=-1
Because the gate can also discard low-frequency target energy, we compare it
with the otherwise identical ungated variant. As
Table~\ref{tab:main} shows, selective gating has higher SI-SDR and is less
susceptible to confusing octaves.
\par\endgroup

\subsection{Model architecture}
\begin{figure}[t]
  \centering
  \includegraphics[width=0.92\linewidth]{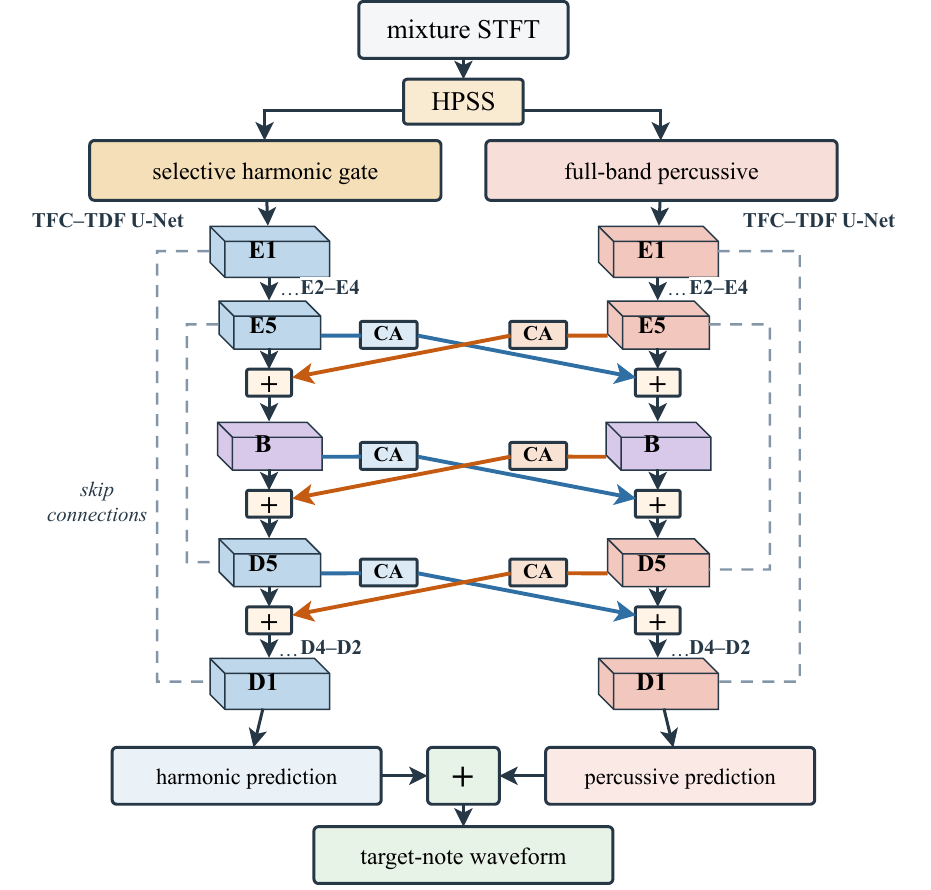}
  \setlength{\abovecaptionskip}{4pt}
  \caption{Architecture of NoteGrab: NoteGrab extracts one
  queried note using gated harmonic and ungated percussive components fed into a two-stream version of TFC--TDF U-Net. At three
  middle stages, a bidirectional cross-attention pair writes information into the opposite
  stream.}
  \label{fig:symmetric_gated_dual_stream}
\end{figure}

Figure~\ref{fig:symmetric_gated_dual_stream} shows the NoteGrab
extractor for one score query.
Each stream uses a 27.4M-parameter TFC--TDF v3 U-Net~\cite{kim2023tfctdf}, which is a model that placed first on Leaderboard B of the 2023 Sound Demixing Challenge.
At the deepest encoder stage, bottleneck, and earliest decoder stage, a bidirectional pair of cross-attention modules reads from one stream and writes to the other in opposite directions. Each
uses one stream as key and value, the other as query, and adds its message to
the queried stream. This is inspired by the cross attention modules used in LadderSym~\cite{chou2026laddersym} for time alignment and ViTTM~\cite{jajal2025vittm} for memory. The resulting model sums the stream outputs in the waveform domain.
Pitch, onset, and offset are used for conditioning via a piano roll like feature map which is applied with FiLM layers at every layer.

\subsection{Joint separation with Adaptive Set Ownership}
\label{ASO}
\begin{figure}[t]
  \centering
  \includegraphics[width=0.92\linewidth]{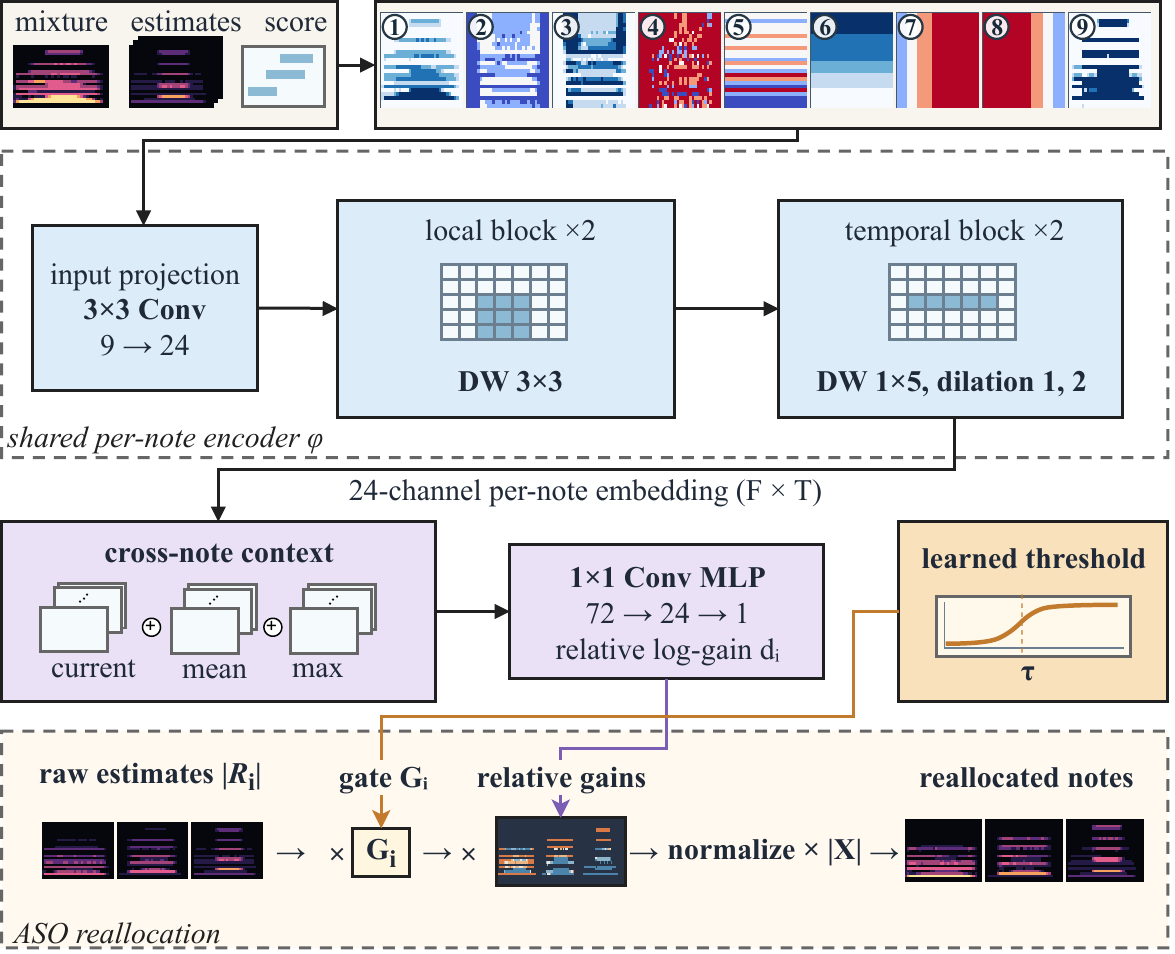}
  \setlength{\abovecaptionskip}{4pt}
  \caption{ASO gates the raw magnitudes,
  predicts relative gains from cross-note context, and normalizes the adjusted
  weights before reallocating mixture energy. 
  The input channels are described in Sec.~\ref{ASO}.}
  \label{fig:aso_architecture}
\end{figure}
Because NoteGrab extracts each note independently, the sum of its outputs do adhere closely to the original mixture. 
ASO is motivated by the idea that joint separation exposes the model to the
other notes, providing context to correct an estimate that may have under- or over-allocated energy. ASO processes 6 seconds of mixture audio and all the notes present within that window. The input to the ASO model uses nine feature maps per note, shown in
Fig.~\ref{fig:aso_architecture}: (1) mixture magnitude; (2) estimate-to-mixture
magnitude ratio; (3) the estimate's share of the summed estimates; (4) per-bin
phase similarity with the mixture; (5) distance to the nearest scored
harmonic; (6) its normalized harmonic index; (7--8) signed onset and acoustic-
offset distances; and (9) a learned gate applied to the input .
The gate is described below: For each raw estimate, \(M_i\) is the magnitude below which the weakest
time--frequency coefficients account for 1\% of its total STFT energy. A learned
global scalar sets \(\tau_i=e^aM_i\). At each bin, ASO applies the two thresholds
that define the gate \(G_i(f,t)\): bins below \(\tau_i/2\) are removed, bins above
\(\tau_i\) are retained, and bins between the thresholds are attenuated using a
cubic function.

Next, a shared \(3\!\times\!3\)
projection maps these features to 24 channels, followed by two local blocks
and two temporal blocks with dilations one and two.
\begingroup\tolerance=9999\emergencystretch=2em\looseness=-1
ASO then concatenates the hidden current maps with mean and max pooled maps over the other concurrent notes to form 72 channels. The network's final prediction is a log-gain $d_i$, yielding adjusted magnitude $w_i = |R_i|G_i e^{d_i}$ and allocation ratio $A_i = w_i / \sum_k w_k$. The estimate is $\widehat S_i = A_i X$ for mixture STFT $X$. If all weights in a bin are zero, ASO falls back to raw note magnitudes with mixture phase to avoid degrading the result by amplifying leakage. This trades off perfect mixture consistency for safer note estimates with less leakage from other notes. The final ASO output is a magnitude mask, as even though complex-mask outputs yielded better independent estimates in NoteGrab, magnitude reallocation simplified mixture consistency, was easier to train and achieved superior performance.
\par\endgroup

\begingroup\tolerance=9999\emergencystretch=4em\looseness=-1
Extracting one note requires one NoteGrab pass. NoteSep splits
a window with $K$ notes requiring $K$ shared-weight NoteGrab passes (batchable)
and one ASO pass for each the individual estimates. When the purpose is note extraction of a single note, incorporating ASO to improve the estimate requires labels for overlapping notes and more than K times the compute. In practice, most applications  already require every note to be extracted (such as editing), so the additional cost of incorporating ASO is dominated by a relatively small constant cost from the ASO network. We use these $K+1$ passes as the inference budget; On an NVIDIA A10, the separator/ASO stages for a 2-s, 16-query input is 1323.8/\allowbreak14.0~GMACs, 454.7/\allowbreak32.7~ms, and
840.1/\allowbreak828.5~MiB peak allocation.
\par\endgroup

\section{Experiments}

\subsection{Experimental setup}
\label{sec:experimental-setup}

\noindent\textbf{Data.}
SCNS-Train contains 25,729 source-complete mixtures and 743,920 isolated-note
targets rendered from six symbolic datasets and NSynth/TinySOL
recorded-note libraries~\cite{manilow2019slakh,hawthorne2019maestro,
wu2022cocochorales,li2019urmp,emiya2010maps,xi2018guitarset,
engel2017nsynth,cella2020tinysol}. SCNS-Eval contains 144 disjoint PDMX scores,
3,473 scored notes, 16 held-out instruments, and held-out note libraries; no
score or audio library is shared with training.

\noindent\textbf{Baselines.}
We compare against Melodyne 5.4.2 and score-informed NMF~\cite{driedger2013scoredecomp,ewert2012scoreinformednmf},
enhanced with inharmonic templates and ADMM~\cite{boyd_distributed_2011}.

\noindent\textbf{Training.}
We apply gain, detuning, pitch shifts, echoes, RIRs, and colored noise; We supervise both streams and their sum
with multiresolution relative complex-STFT, log-magnitude, waveform, and
log-envelope losses. The NoteGrab variants use 16~kHz
audio, 1,024-sample STFTs with 128-sample hops, and 30,000 steps (batch
8, AdamW, learning rate \(5\times10^{-5}\), cosine decay); For simplicity, ASO uses the same recipe.

\noindent\textbf{Metrics.}
We report SI-SDR~\cite{leroux2019sdr}, off-target energy, and target projection gain;
SI-SDR is per-note median (mean), target projection gain should approach one, and off-target
energy is the target-normalized residual after projection. Octave confusion is the
fraction of target events with an active octave competitor at \(p_i\!\pm\!12\)
semitones for which the competitor has higher
\(\rho^2(e,s)=\langle e,s\rangle^2/(\lVert e\rVert_2^2\lVert s\rVert_2^2)\)
than the target, where \(e\) is an estimate and \(s\) its reference.

\subsection{Note separation results}

\begin{table}[t]
\centering
\setlength{\abovecaptionskip}{4pt}
\caption{SCNS-Eval results. SI-SDR is per-note median (mean) in dB; Off-E and
gain are medians. Higher SI-SDR is better; lower Off-E and Oct-Conf are better;
gain should approach one. $^\dagger$Melodyne uses matched notes only.
}
\label{tab:main}
\small
\setlength{\tabcolsep}{1.5pt}
\renewcommand{\arraystretch}{0.85}
\begin{tabular}{@{}lrrrr@{}}
\toprule
\textbf{System} & \textbf{SI-SDR} & \textbf{Off-E} & \textbf{Gain} & \textbf{Oct-Conf $\downarrow$}\\
\midrule
Melodyne 5.4.2$^\dagger$ & -3.88 (-4.86) & 0.62 & 0.34 & --\\
Score-Informed NMF & 2.49 (2.27) & 0.25 & 0.82 & --\\
Independent Ungated & 3.94 (3.59) & \textbf{0.09} & 0.62 & 8.15\%\\
Independent Selective & 4.46 (4.24) & 0.09 & 0.62 & 7.79\%\\
NoteGrab Ungated & 4.54 (4.28) & 0.11 & 0.71 & 8.29\%\\
NoteGrab Selective & 5.09 (4.87) & 0.12 & 0.72 & 5.19\%\\
NoteSep & \textbf{7.39 (7.35)} & 0.11 & \textbf{0.83} & \textbf{2.96\%}\\
\bottomrule
\end{tabular}
\renewcommand{\arraystretch}{1.0}
\end{table}

\begingroup\tolerance=9999\emergencystretch=2em\looseness=-1
Melodyne does not use the score and obtains notes by clustering components detected in the mixture, so it is not directly comparable with the score-informed methods. For completeness, we match its detections to score events using pitch and onset alone. It matches 1,890 of 3,473 scored notes (54.42\%) at the correct pitch and within 150~ms, and reaches a median SI-SDR of $-3.88$~dB on the matched outputs, compared with 9.24~dB for NoteSep. The demo provides qualitative comparisons. In the paired ablations, bidirectional cross-attention (NoteGrab) and selective gating each improve SI-SDR. Adding ASO yields the strongest overall improvement, increasing SI-SDR to 7.39~dB while also reducing off-target energy relative to NoteGrab Selective.
\par\endgroup
\begin{figure}[t]
  \centering
  \includegraphics[width=0.82\linewidth]{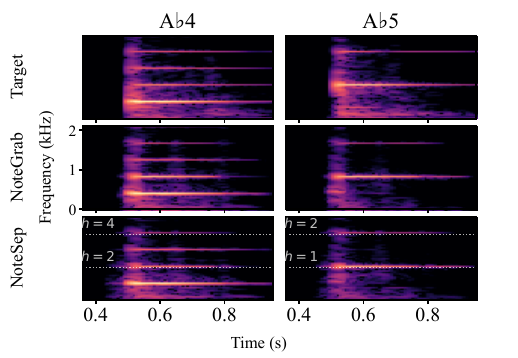}
  \setlength{\abovecaptionskip}{4pt}
  \caption{Octave-overlapping A$\flat$4 and A$\flat$5 notes in an SCNS-Eval
piano chord. From top: isolated target, independent NoteGrab estimate, and
joint NoteSep output. Dotted lines mark shared partials at 0.83 and
1.66~kHz; \(h\) labels denote harmonic number.}
  \label{fig:qualitative}
\end{figure}

\begingroup\tolerance=9999\emergencystretch=2em\looseness=-1
Figure~\ref{fig:qualitative} shows how this joint allocation changes one
octave-overlapping chord. Near 831~Hz, A$\flat$4's second harmonic coincides
with the A$\flat$5 fundamental, and their combined raw claim exceeds the
available mixture energy. ASO lowers both estimates. Near 1.66~kHz, it moves energy
from A$\flat$4's fourth harmonic toward A$\flat$5's second harmonic. The largest
overall correction is for A$\flat$4, whose SI-SDR rises 5.16~dB.
\par\endgroup



\begin{table}[t]
\centering
\setlength{\abovecaptionskip}{4pt}
\caption{SCNS-Eval pre-edit SI-SDR (as in Table~\ref{tab:main}) and 13,892
subsequent edits. Values are medians. Higher is better.}
\label{tab:editing}
\small
\setlength{\tabcolsep}{2.2pt}
\renewcommand{\arraystretch}{0.85}
\begin{tabular}{@{}lrrrr@{}}
\toprule
System & Pre-edit & Target 2f & Unedited 2f & Final 2f \\
\midrule
Score-Informed NMF & 2.49 & 25.73 & 41.77 & 43.40 \\
NoteGrab & 5.09 & 31.70 & 32.87 & 32.38 \\
NoteSep & \textbf{7.39} & \textbf{40.09} & \textbf{71.23} & \textbf{71.08} \\
\bottomrule
\end{tabular}
\renewcommand{\arraystretch}{1.0}
\end{table}


\subsection{Editing evaluation}

\begingroup\tolerance=9999\emergencystretch=2em\looseness=-1
NoteSep can be applied to structural editing of music (pitch, timing). We evaluate on SCNS-Edit, which is SCNS-Eval with edits applied to the ground truth notes. For each edit, we compare three signals with their ground-truth counterparts:
the transformed target note, the sum of unedited note estimates, and their
final sum. This separates errors in the requested edit from errors in preserving
the rest of the performance. Each signal is formed from stored note estimates;
we do not subtract an estimate from the input mixture. We test rebalancing,
retiming, retuning, and repitching, giving 13,892 edits in total. SI-SDR is only used for evaluating the separated note pre-transformation. To evaluate quality after edits are applied and remixed back into the mixture post transformation, we use the 2f metric~\cite{torcoli2021objective}, which passes the edited reference and estimate through
the PEAQ auditory model. As shown in Table~\ref{tab:editing}, NoteGrab improves pre-edit separation and edited-target quality over Score-Informed NMF, but independently extracted notes from NoteGrab have worse performance when summed due to the unenforced mixture consistency. NoteSep mitigates this inconsistency with ASO, substantially improving the unedited and final-sum 2f scores while achieving the best performance across all metrics.
\par\endgroup

\subsection{Instrument separation}

\begin{table}[t]
\centering
\setlength{\abovecaptionskip}{4pt}
\caption{Instrument separation in dB: median and mean.}
\label{tab:zeroshot}
\small
\setlength{\tabcolsep}{1.2pt}
\renewcommand{\arraystretch}{0.72}
\begin{tabular}{@{}lS[table-format=-1.2]S[table-format=-1.2]S[table-format=2.2]S[table-format=2.2]S[table-format=-1.2]S[table-format=-1.2]@{}}
\toprule
\textbf{System} & \multicolumn{2}{c}{\textbf{SI-SDR $\uparrow$}} & \multicolumn{2}{c}{\textbf{SI-SDRi $\uparrow$}} & \multicolumn{2}{c}{\textbf{SDR $\uparrow$}}\\
 & {Med.} & {Mean} & {Med.} & {Mean} & {Med.} & {Mean}\\
\midrule
\multicolumn{7}{@{}l}{\textit{PHENICX-Anechoic, 4 pieces}}\\
X-UMX & -4.78 & -7.69 & 4.25 & 3.33 & 0.76 & {--}\\
Score-Inf. X-UMX & -4.79 & -6.52 & 5.12 & 4.50 & 1.04 & {--}\\
NoteGrab & -4.04 & -4.69 & 6.09 & 6.33 & -1.32 & {--}\\
NoteSep & \multicolumn{1}{r}{\textbf{$-1.64$}} & \multicolumn{1}{r}{\textbf{$-2.04$}} & \multicolumn{1}{r}{\textbf{8.45}} & \multicolumn{1}{r}{\textbf{8.98}} & \multicolumn{1}{r}{\textbf{1.53}} & {--}\\
\midrule
\multicolumn{7}{@{}l}{\textit{Bach10, 10 pieces}}\\
Miron Score-Inf. CNN & \multicolumn{1}{r}{\textbf{6.17}} & \multicolumn{1}{r}{\textbf{5.52}} & \multicolumn{1}{r}{\textbf{10.73}} & \multicolumn{1}{r}{\textbf{10.28}} & 6.76 & 6.84\\
Miron Score-Inf. NMF & 5.12 & 4.54 & 9.81 & 9.31 & 5.61 & 5.80\\
NoteGrab & 4.12 & 3.20 & 8.75 & 7.96 & 5.73 & 5.22\\
NoteSep & 5.31 & 4.86 & 10.10 & 9.62 & \multicolumn{1}{r}{\textbf{6.83}} & \multicolumn{1}{r}{\textbf{6.47}}\\
\bottomrule
\end{tabular}
\renewcommand{\arraystretch}{1.0}
\end{table}

\begingroup\tolerance=9999\emergencystretch=2em\looseness=-1
The purpose of this evaluation is to test transfer to real recordings. We do so
through score-informed instrument separation, aggregating note estimates by
instrument over four PHENICX-Anechoic pieces (38 stems; 11,410 note queries)
and ten Bach10 quartets (40 stems). NoteGrab and NoteSep are trained only on
SCNS-Train and see neither dataset during training. We use released X-UMX and
score-informed X-UMX models on PHENICX, and released score-informed CNN and
NMF results on Bach10~\cite{tunturi2025scoremss,miron2016orchestral,miron2017monaural}.
Miron CNN was trained on
synthetic renditions of the evaluated Bach10 scores.
\par\endgroup

\begin{samepage}
\begingroup\tolerance=9999\emergencystretch=2em\looseness=-2
As shown in Table~\ref{tab:zeroshot}, NoteSep improves over NoteGrab on both
datasets. On PHENICX, SI-SDRi rises from 6.09 to 8.45~dB and SDR from $-1.32$
to 1.53~dB; on Bach10, SI-SDRi rises from 8.75 to 10.10~dB and SDR from 5.73
to 6.83~dB. NoteSep remains slightly below the task-specific Miron CNN in SI-SDRi but outperforms the Miron NMF model. The demo contains listenable results.
\par\endgroup
\end{samepage}

\section{Conclusion}

\begingroup\tolerance=9999\emergencystretch=3em\looseness=-3
This paper presents the first method for deep learning-based note extraction and separation. We present the datasets SCNS-Train containing 743,920 training targets, and
SCNS-Eval that provides a held-out 16-instrument benchmark. On SCNS-Eval, NoteSep reaches 7.39~dB median
SI-SDR, compared with 2.49~dB for Score-Informed NMF. The accompanying artifact provides code, datasets, model checkpoints, audio examples, and an interactive note-level visualization. NoteSep presents new capabilities to potentially impact many application spaces in music such as music editing, musical-line highlighting,
music performance analysis, music information retrieval tasks like transcription and music education.
\par\endgroup

\bibliographystyle{IEEEbib}
\bibliography{refs,refs_extra}

\end{document}